\documentclass[%
 reprint,
 amsmath,amssymb,
 aps,
]{revtex4-2}
\usepackage{graphicx}
\usepackage{dcolumn}
\usepackage{bm}
\usepackage{xcolor}
\usepackage{hyperref}

\usepackage{braket}
\begin{document}

\preprint{APS/123-QED}

\title{Role of CISS in the Radical Pair Model of Avian Magnetoreception 
}

\author{Yash Tiwari}

\author{Vishvendra Singh Poonia}%
 \email{vishvendra@ece.iitr.ac.in}
\affiliation{%
Department of Electronics and Communication, Indian Institute of Technology, Roorkee, Uttrakhand 247667, India 
}%




\date{\today}

\begin{abstract}
In this paper, we investigate the effect of chiral-induced spin selectivity (CISS) on the radical pair model of avian magnetoreception. We examine the impact of spin selectivity on the avian compass sensitivity. In this analysis, we also consider the dipolar and exchange interactions and observe their interplay with CISS. We find that CISS results in multifold increase in avian compass sensitivity. Interestingly, we also observe that CISS can counter the deleterious effect of dipolar interaction and increase system sensitivity.
The analysis has been performed for both the toy model (only 1 nucleus) and a more general case where we consider 4-nuclei from the cryptochrome radical pair system. We observe that the CISS allows the radical pair model to have a more realistic rate with good sensitivity. We also do analysis of `functional window' of avian compass reported in  behavioral experiments in functional window. We could not find a parameter set where `functional window' can be observed along with CISS. 


\end{abstract}

\maketitle


\section{\label{INTRODUCTION}Introduction}

The avian magneto-reception is the ability of migratory birds to sense the geomagnetic field and navigate with its assistance. Two alternative theories are proposed to understand the avian magnetoreception. One is based on magnetite~\cite{kirschvink2001magnetite,fleissner2007novel} and the other on the radical pair model~\cite{ritz2000model}  with evidence strongly favoring the later~\cite{schulten1978biomagnetic,ritz2000model,hore2016radical}.

The radical pair model (RPM) is based on the spin of two electrons created on adjacent radicals (donor and acceptor). The formation of this radical pair is attributed to the photo-excitation of the donor and acceptor molecules whereby an electron transfer is involved. The photo-excitation happens due to light of a particular frequency falling on the bird's retina, which is the site of these donor and acceptor molecules. The spin of these two electrons interacts with the Earth's magnetic field and hyperfine field due to surrounding nuclei. In addition, there is electron-electron interaction in the form of dipolar and exchange interactions. All these interactions affect the final entity that is obtained after recombination. Evidences suggest that the donor and acceptor molecules are co-factors of cryptochrome molecule ~\cite{ritz2000model,liedvogel2007chemical,maeda2012magnetically,zoltowski2019chemical,xu2021magnetic}.

However, an important aspect, often disregarded, is the electron transfer medium and how it might affect the system. The electron transfer between donor and acceptor essentially happens in a chiral medium. 
And according to chiral-induced spin selectivity (CISS) effect, electron transport in a chiral medium is spin selective~\cite{ray1999asymmetric} which illustrates that chiral molecules act as a spin filter and a certain chirality only allows electron of a particular spin to travel through it~\cite{gohler2011spin,naaman2012chiral,naaman2015spintronics}. The reason attributed to this spin selective assistance is the spin-orbit interaction, which effectively interacts with the linear momentum of the electron. The electrostatic potential provided by the chiral molecules' geometric structure accelerates the electron's momentum having a particular kind of spin~\cite{dalum2019theory,michaeli2019origin,matityahu2016spin}. Recently, Luo et al. analyzed the radical pair model of avian compass with CISS~\cite{luo2021chiral}. This calls for more comprehensive analysis of effect of CISS on the radical pair spin dynamics and various behavioral characteristics of the avian compass.

In this work, our focus is to understand interplay between CISS and various paramters of the radical pair model of avian compass and we analyze the `functional window' characteristics of the avian compass in the light of CISS. `Functional window' is the behavioral characteristics of the compass that refers to the selectivity of the compass around the geomagnetic fields (25$\mu$T to 65 $\mu$T)~\cite{wiltschko1972magnetic,poonia2017functional,winklhofer2013avian,walcott1978animal,10.1111/j.1365-246X.2010.04804.x}. This is an important yet not well-understood feature of the avian compass especially from the point of view of the radical pair model.  
We study the effect of CISS for various recombination rates of radicals. Additionally, dipolar and exchange interactions are usually detrimental to the action of the avian compass and cause reduction in the compass sensitivity. However, it was observed that exchange and dipolar interaction can partially cancel each other, thereby restoring the sensitivity to some extent\cite{efimova2008role}. It becomes important to observe how electron-electron interaction affects sensitivity in conjunction with CISS.
We begin the work by observing the effect of CISS on system yield, sensitivity, and functional window for a radical pair toy model (where only one nucleus is considered) and a realistic cryptochrome based radical pair system where we consider two nuclei each on flavin adenine dinucleotide FAD (two nitrogen nuclei) and tryptophan TrpH (one nitrogen and one hydrogen) radicals. The FAD act as a donor entity, whereas TrpH act as an acceptor entity. Following this, we also observe the effect of dipolar and exchange interaction on the CISS-assisted magnetoreception. 

The manuscript has been organized as follows: Section~\ref{METHODOLOGY} discusses the simulation methodology followed for analysis. Section ~\ref{RESULTS}  discusses the results, where subsection~\ref{CISSYIELDSENSTIVTY} discusses the effect of CISS on sensitivity and  subsection~\ref{CISSYIELDWINDOW} discusses the effect of CISS on the functional window. Subsection~\ref{ELECTRONELECTRONINTERACTION} explores the effect of electron-electron interaction on sensitivity along with CISS. All calculations were performed using MATLAB 2021.

\section{\label{METHODOLOGY}Methodology}
\begin{figure}[htbp]
\centering
\includegraphics[width=90mm,keepaspectratio]{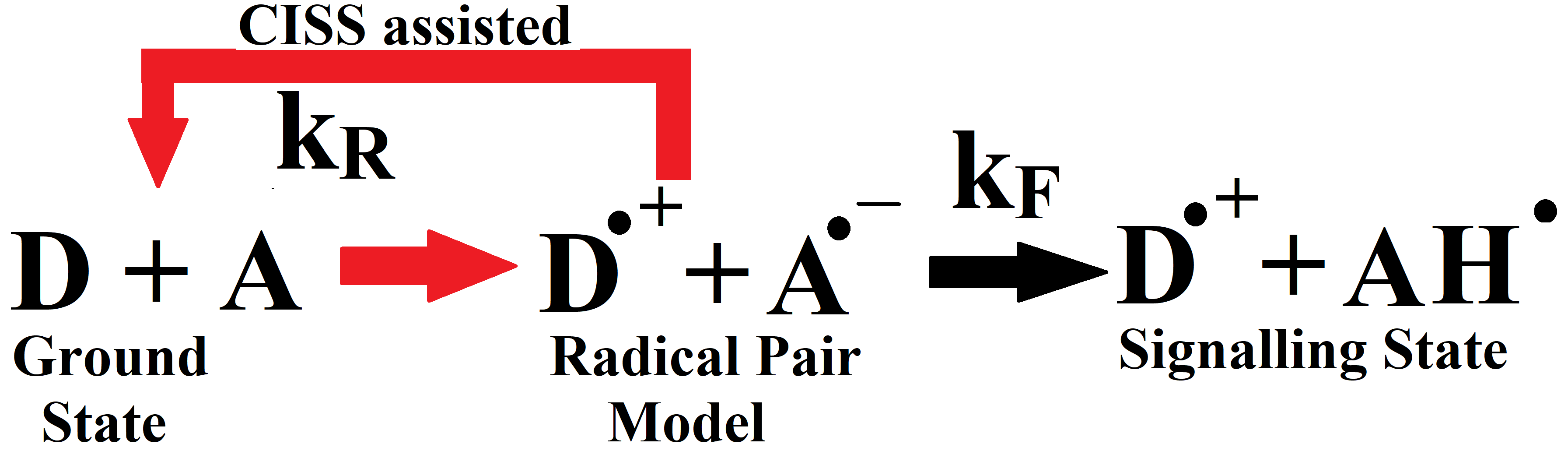}
\caption{Schematic of the radical pair mechanism with CISS where D denotes the donor molecule and A represents the acceptor molecule. The red curve correspond to CISS assisted recombination to the ground state.}
\label{Fig1_RPMechanism}
\end{figure}

We study the cryptochrome (four nuclei) based radical pair model of avian compass ~\cite{xu2021magnetic,einwich2022localisation,wong2021cryptochrome}.
In this model, photoexcitation of acceptor molecule leads electron from ground to excited state leaving vacancy in the ground state. An adjacent donor molecule donates an electron to fill the vacancy in the ground state of the acceptor. 
Hence a radical pair is formed: one unpaired electron in the ground state of the donor (spin operator $S_D$) and the other unpaired electron in the excited state of the acceptor (spin operator $S_A$). The radical pair recombines back to form the `ground' state of the system or forms a `signaling' state after protonation of the acceptor molecule. The schematic of this mechanism is given in Fig.~\ref{Fig1_RPMechanism}. 
We account for CISS in the formation of radical pair as a result of electron transfer from donor to acceptor and in formation of ground state due to reverse electron transfer. However, formation of signaling state from the radical pair does not involve any electron transfer, therefore, CISS does not play any role in this leg of RPM.
The spin selectivity of the system due to CISS (shown in red arrow in Fig.~\ref{Fig1_RPMechanism}) is captured by the initial state and the ground state formed after recombination. 
This model is discussed in more detail in Ref.\cite{luo2021chiral}.

The spin state of the radical pair can be singlet state or triplet state or a superposition thereof. The spin dynamics is governed by the Hamiltonian:~\cite{cintolesi2003anisotropic,fay2020quantum,luo2021chiral}
\begin{equation} 
\label{Hamiltonian_RPM}
\begin{split}
\hat{H}  = \omega.(\hat{S}_{A}+\hat{S}_{D}) + \sum_{i\in{D,A}}\sum_{k}\hat{S}_{i}.A_{ik}.\hat{I}_{ik} \\ -
J(2\hat{S}_{A}.\hat{S}_{D} + 0.5) +
\hat{S}_{A}.D.\hat{S}_{B}
\end{split}
\end{equation}

$\hat{S}_{A}$ and $\hat{S}_{D}$ are spin of electron on donor and acceptor molecule, $\omega=g\Bar{\mu_B} \Bar{B}$ where $\Bar{B} = B_0{((cos\theta cos\phi)\Bar{x}+(cos\theta sin\phi)\Bar{y}+(cos\theta)\Bar{z}})$. $B_0$ corresponds to the earth's magnetic field, which in our works is assumed to be 50$\mu$T. $\theta$ and $\phi$ describe orientation of magnetic field with respect to hyperfine tensor.\cite{gauger2011sustained} . $J$ is the electron-electron spin-exchange interaction, whereas D is the dipole-dipole interaction tensor. The form of dipolar Hamiltonian depends upon the relative direction of electron spin with each other and the externally applied magnetic field. $A_{ik}$ is the hyperfine interaction between electron and nuclear spin x,y,z, components. 
 
The initial density of matrix of system is given by $P_I= \ket{\psi_I}\bra{\psi_I}\otimes \frac{I}{Z}$, where $\frac{I}{Z}$ corresponds to the normalized mixed state of the nuclei (Z is sized of combined hilbert space of nuclei). $\ket{\psi_I}$ is the initial state of the radical pair. If the CISS-assisted recombination back to the ground state is represented by $\ket{\psi_R}$, then: 

\begin{equation} 
\label{INITIAL_STATE}
\begin{split}
\ket{\psi_I}=\frac{1}{\sqrt{2}}[\sin(0.5\chi)+\cos(0.5\chi)]\ket{\uparrow_D\downarrow_A}+\\
\frac{1}{\sqrt{2}}[\sin(0.5\chi)-\cos(0.5\chi)]\ket{\downarrow_D\uparrow_A}
\end{split}
\end{equation}
\begin{equation} 
\label{RECOMBINATION_STATE}
\begin{split}
\ket{\psi_R}=-\frac{1}{\sqrt{2}}[\sin(0.5\chi)-\cos(0.5\chi)]\ket{\uparrow_D\downarrow_A}-\\
\frac{1}{\sqrt{2}}[\sin(0.5\chi)+\cos(0.5\chi)]\ket{\downarrow_D\uparrow_A}
\end{split}
\end{equation}

The parameter $\chi\in [0,\frac{\pi}{2}]$ depends on the spin selectivity of the medium in which the reaction is taking place. When $\chi=0$,
$\ket{\psi_I}=\frac{1}{\sqrt{2}}\ket{\uparrow_D\downarrow_A}-\ket{\downarrow_D\uparrow_A}$ and $\ket{\psi_R}=\frac{1}{\sqrt{2}}\ket{\uparrow_D\downarrow_A}-\ket{\downarrow_D\uparrow_A}$ but when $\chi=\frac{\pi}{2}$ ,
$\ket{\psi_I}=\ket{\uparrow_D\downarrow_A}$ and $\ket{\psi_R}=\ket{\downarrow_D\uparrow_A}$. These depict the two extrema of the spin selectivity of the chiral medium. When $\chi=0$, it depicts the case when no spin selectivity is introduced in the medium (conventional RPM), whereas $\chi=\frac{\pi}{2}$ depicts the case when the medium is fully chiral, showing a full medium selectivity. All the other cases are intermediate cases.
Fig.\ref{Fig1_RPMechanism} depicts the CISS assisted chemical reaction where $k_R$ is rate of recombination of radical back to ground state. $k_F$ is rate at which the radical pair protonate with $H^{+}$ to create signaling state.
\begin{equation} 
\label{MASTER_EQUATION}
\begin{split}
{\frac{d\hat{\rho}}{dt}=-(Coherent  + Recombination.)}\\
=-i[\hat{H},\hat{\rho}(t)]-\frac{1}{2}k_R[\ket{\psi_R}\bra{{\psi_R}},\hat{\rho}(t)]-k_F\hat{\rho}(t)
\end{split}
\end{equation}
The master equation governing the system is given by Eq.\ref{MASTER_EQUATION} where the $Coh.$ correspond to the coherent evolution of the system. $Recomb.$ corresponds to the chemical dissipation responsible for the formation of yield product which enales birds to do magnetoreception. In Liouville space, this is accomplished by using a projection operator, which maps the yield of the system into the shelving states as done in \cite{gauger2011sustained,poonia2015state}

\section{\label{RESULTS}Results and Discussion}
This section is divided into three subsections. In the first two subsections, we report the effect of CISS on sensitivity and functional window in the absence of electron-electron interactions (exchange and dipolar interactions). In the last subsection, we also include dipolar and exchange interactions in CISS framework and observe the sensitivity of the compass.

\subsection{\label{CISSYIELDSENSTIVTY}Effect of CISS on Senstivity}
To define the sensitivity of the avian compass, the yield product of either signaling state ($\phi_F$) or ground state ($\phi_R$) state needs to be defined. The yield of the signaling state ($\phi_F$) is defined by Eq.~\ref{YIELD_PARTIALSTATE} where $\hat{\rho(t)}$ is the solution of the master equation Eq.~\ref{MASTER_EQUATION}, $Tr$ is the trace over the state density matrix $\rho$. $k_F$ is the rate associated with the signaling state. The yield of ground state ($\phi_R$) associated with recombination operator $\ket{\psi_R}\bra{{\psi_R}}$ can easily be found by $\phi_R$ + $\phi_F$ =1. We use the semi-classical formalism of Eq.~\ref{YIELD_PARTIALSTATE} to avoid solving in Liouville Space which poses a significant computation challenge. The formalism has been initially proposed in \cite{schulten1978semiclassical} and reiterated in \cite{hiscock2018long}     
\begin{equation} 
\label{YIELD_PARTIALSTATE}
\begin{split}
\phi_F=k_F\int_{0}^{\infty} P_S(t)dt=k_F\int_{0}^{\infty} Tr[\hat{\rho(t)}]dt
\end{split}
\end{equation}

\begin{equation} 
\label{SenstivityExp}
\begin{split}
\triangle\phi_R=\underset{\theta,\phi}{max}(\phi_R)-\underset{\theta,\phi}{min}(\phi_R)
\end{split}
\end{equation}
The sensitivity of the avian compass is defined in Eq.\ref{SenstivityExp}. The difference between the maximum and minimum $\phi_R$ is taken all over the values of $\phi$ and $\theta$.
A measure of sensitivity can be either $\triangle\phi_R$ or $\triangle\phi_F$. In our work, we choose $\triangle\phi_R$ to describe the sensitivity. 
Sensitivity quantifies the yield range for the same $\theta$ and $\phi$, which essentially describes the longitudinal and latitudinal position on earth. We ideally require this sensitivity to be as high as possible for better operation of avian compass.

\begin{figure}[htbp]
\centering
\includegraphics[width=90mm,keepaspectratio]{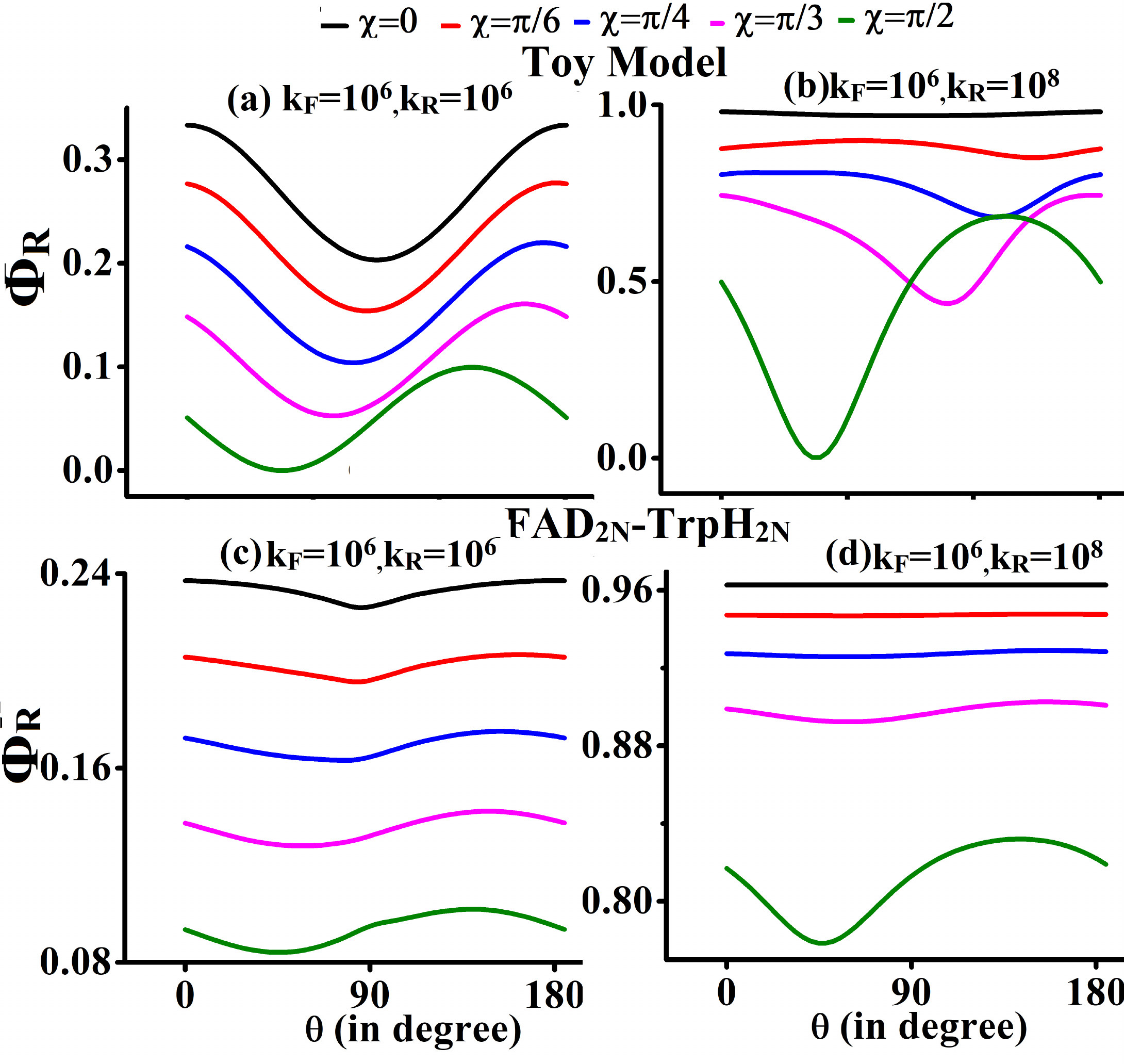}
\caption{Yield of ground state ($\phi_R$) vs $\theta$ for constant $\phi=0$ for five distinct value of $\chi$ (showing varying degree of spin selectivity due to CISS. $0^o$(black), $30^o$(red), $45^o$(blue), $60^o$(pink), $90^o$(green). This is shown at two values of signalling state recombination rate $(k_F)$ and ground state recombination rate $(k_R)$. (a)$k_F = 10^6 s^{-1}, k_R = 10^6 s^{-1}$, (b)$k_F = 10^6 s^{-1}, k_R = 10^8 s^{-1}$. (a) and (b) in top row show the yield for a toy model at the two specified rates for specified value of $\chi$. (c) and (d) in bottom row show the yield for the specified cryptochrome based four-nuclei RP $(FAD_{2N}-TrpH_{2N})$ system at the two rates for various values of $\chi$.}
\label{YIELD_VS_THETA}
\end{figure}

The toy model we consider consist of two electron and  a single nuclei. One electron interacts with the nuclei having spin $\frac{1}{2}$. The cryptochrome (four nuclei) system is modelled with each electron having hyperfine interaction with two nuclei.
Fig.~\ref{YIELD_VS_THETA} shows the product yield of the ground state ($\phi_R$) as a function of $\theta$ for varying degrees of spin selectivity due to CISS.
Each color corresponds to the product yield at a single value of $\chi$ depicting a certain degree of spin selectivity for $\phi=0$ and $\theta\in\lbrace 0^o, 180^o\rbrace$. Fig.~\ref{YIELD_VS_THETA} (a) and (b) correspond to $\phi_R$ vs. $\theta$ when $k_F=10^6s^{-1},k_R=10^6s^{-1}$ and $k_F=10^6 s^{-1},k_R=10^8 s^{-1}$ respectively for the toy model. Fig.~\ref{YIELD_VS_THETA}.(c),(d) correspond to case $k_F=10^6 s^{-1},k_R=10^6 s^{-1}$ and $k_F=10^6 s^{-1},k_R=10^8 s^{-1}$ respectively for the  for the realistic case of cryptochrome (four nuclei) based RP system.
We observe from Fig.~\ref{YIELD_VS_THETA} that the yield plot is symmetric about $\theta=90^o$ when $\chi=0$, but such is not the case when $\chi=\frac{\pi}{2}$. Hence, we lose symmetry of the yield with respect to $\theta$ due to CISS. Interestingly, we also observe that  $\triangle\phi_R$ is dramatically increased for $\chi=$ $\frac{\pi}{2}$ case in Fig.~\ref{YIELD_VS_THETA}.(b) compared to $\chi=0$. $\triangle\phi_R$ is comparable for both of these extreme cases in Fig.~\ref{YIELD_VS_THETA}. (a). We observe this for a single $\phi=0$, assuring us that the minimum value of $\triangle\phi_R$ is quite high for CISS-assisted avian magneto reception into the toy model. A similar observation is made for the four-nuclei RP case also.
From Fig.\ref{YIELD_VS_THETA}, we observe that for $k_F=10^6s^{-1},k_R=10^6s^{-1}$, the sensitivity remain more or less same as we increase CISS, but the yield of ground state decreases (signaling state yield increases). For $k_F=10^6s^{-1},k_R=10^8s^{-1}$, we observe an increase in sensitivity with CISS, however an increase of signaling state yield is observed with CISS. This was observed for both toy model and four nuclei model for cryptochrome system.

We plot sensitivity for rate combination $k_F,k_R\in [10^4,10^9] s^{-1}$, giving us a more holistic picture.
\begin{figure}[htbp]
\centering
\includegraphics[width=90mm,keepaspectratio]{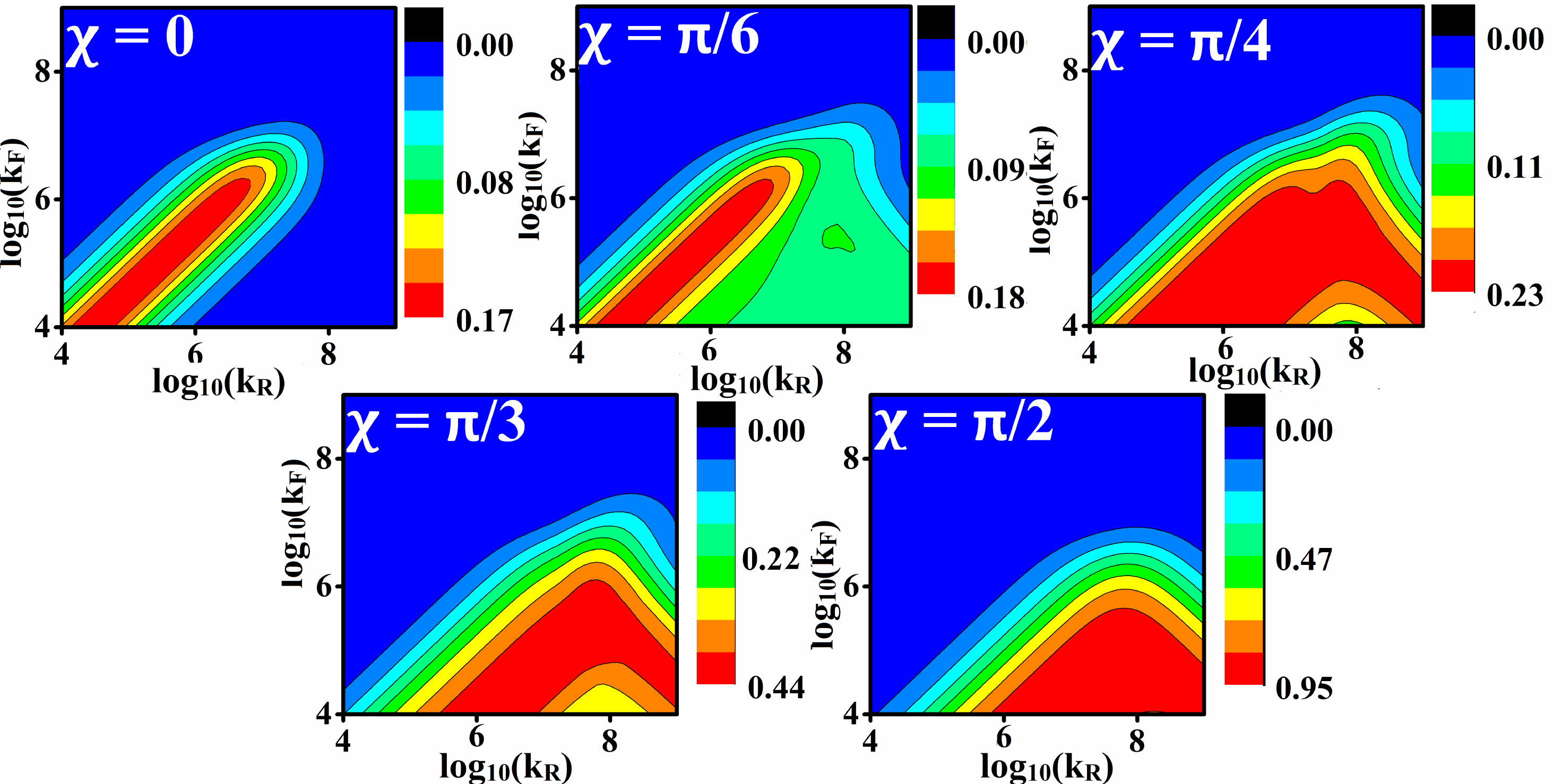}
\caption{Contour plots for sensitivity ($\triangle\phi_R$) as a function of $log_{10}k_F$ and $log_{10}k_R$ for five distinct value of $\chi$ showing varying degree of spin selectivity due to CISS ($0^o$, $30^o$, $45^o$, $60^o$, $90^o$). The plots corresponds to the toy model.}
\label{Contour_Plot}
\end{figure}
We have plotted sensitivity $\triangle\phi_R$ in Fig.\ref{Contour_Plot} for toy model and in Fig.~\ref{Contour_Plot4N} for cryptochrome (four nuclei) RP system. This has been done for various values of rate $k_F,k_R$ for five values of $\chi$. 
\begin{figure}[htbp]
\centering
\includegraphics[width=90mm,keepaspectratio]{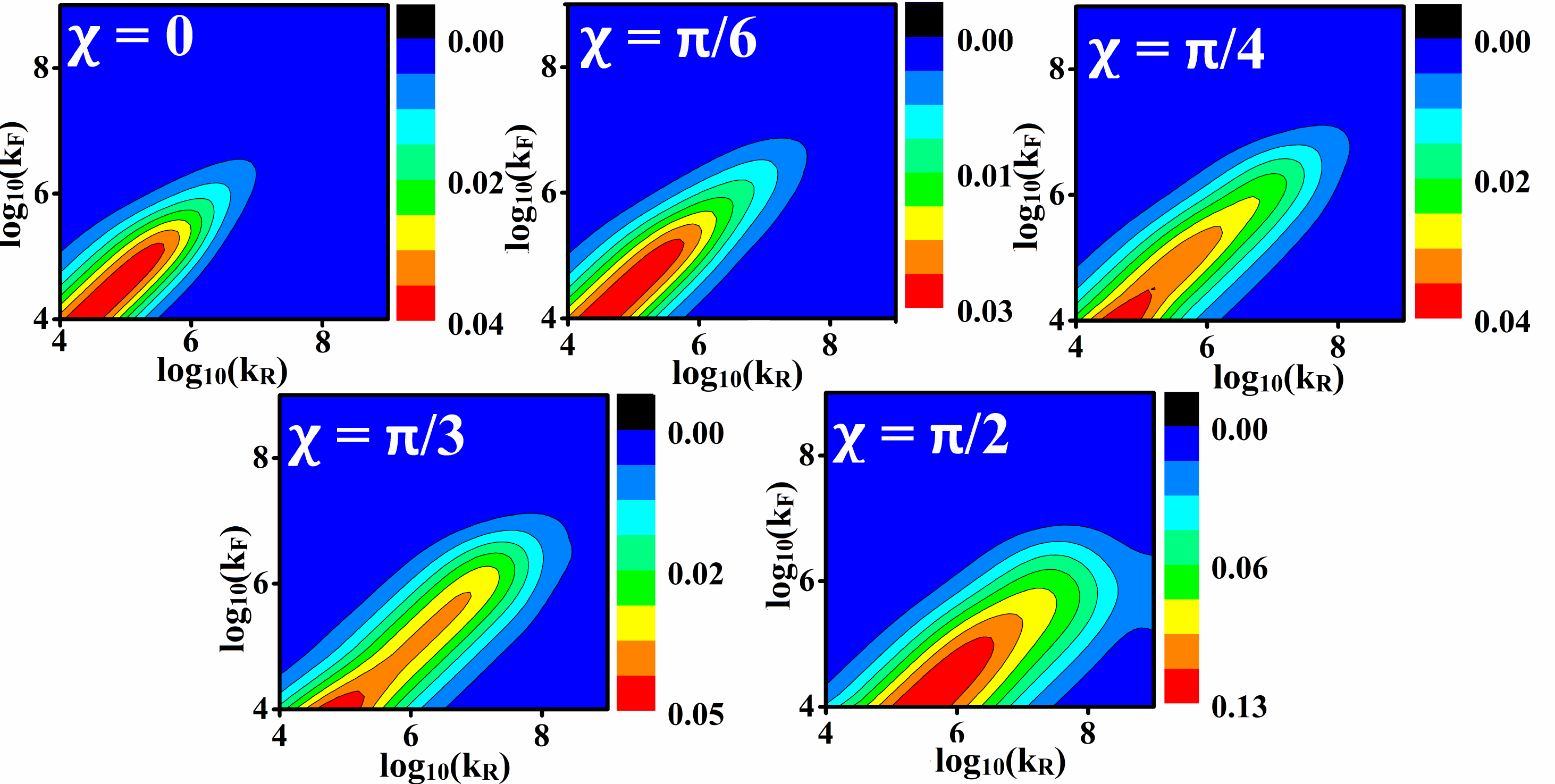}
\caption{Contour plots for sensitivity ($\triangle\phi_R$) as a function of $log_{10}k_F$ and $log_{10}k_R$ for five distinct value of $\chi$ showing varying degree of spin selectivity due to CISS ($0^o$, $30^o$, $45^o$, $60^o$, $90^o$). The plot corresponds to the cryptochrome based RP $(FAD_{2N}-TrpH_{2N})$ system with 4 nuclei.}
\label{Contour_Plot4N}
\end{figure}
As observed in Fig.~\ref{YIELD_VS_THETA}, we observe the similarity of sensitivity between toy and cryptochrome (four nuclei) RP system. We observe in these plots that the maximum sensitivity  $\triangle\phi_R$ is attained at $\chi =\frac{\pi}{2}$. A major difference between the plots of sensitivity of toy and cryptochrome (four nuclei) molecules is the skewness of the area of maximum sensitivity. The rate combination of $k_F,k_R$ for which we can derive sensitivity is far less for the cryptochrome (four nuclei) case than the toy model. Also, at $\chi=\frac{\pi}{6}$, we observe that the increase in sensitivity is very low for the toy model (with respect to $\chi=0$, about 0.01), but it actually decreases for cryptochrome for a similar comparison. After analyzing Fig.~\ref{Contour_Plot} and Fig.~\ref{Contour_Plot4N}, we observe that at high CISS, sensitivity is significant even at high rates of radical recombination. We observe that the protonation rate ($k_F$) is a limiting factor in the RP model. If $k_F$ is more the $10^6 s^{-1}$, the sensitivity becomes insignificant. Important point to note here is that the range of recombination rate to the geround state ($k_R$) is increased in presence of CISS. Thus CISS allows the radical pair model to have more realistic recombination rates (upto $10^8 s^{-1}$). In a nutshell, we clearly demonstrate that CISS is increasing the sensitivity of the compass with more realistic recombination rates.

\subsection{\label{CISSYIELDWINDOW}Effect of CISS on Funcitonal Window}
The functional window is the selectivity of the avain compass with respect to the external magnetic field. It can be studied by analyzing sensitivity ($\triangle\phi_R$) as a function of the magnetic field. the system with respect to the external magnetic field.  Based on the behavioral experiment, we expect that the avian compass show maximum sensitivity around the earth's magnetic field and a reduced sensitivity around other magnetic fields \cite{winklhofer2013avian,ritz2000model}. 

\begin{figure}[htbp]
\centering
\includegraphics[width=90mm,keepaspectratio]{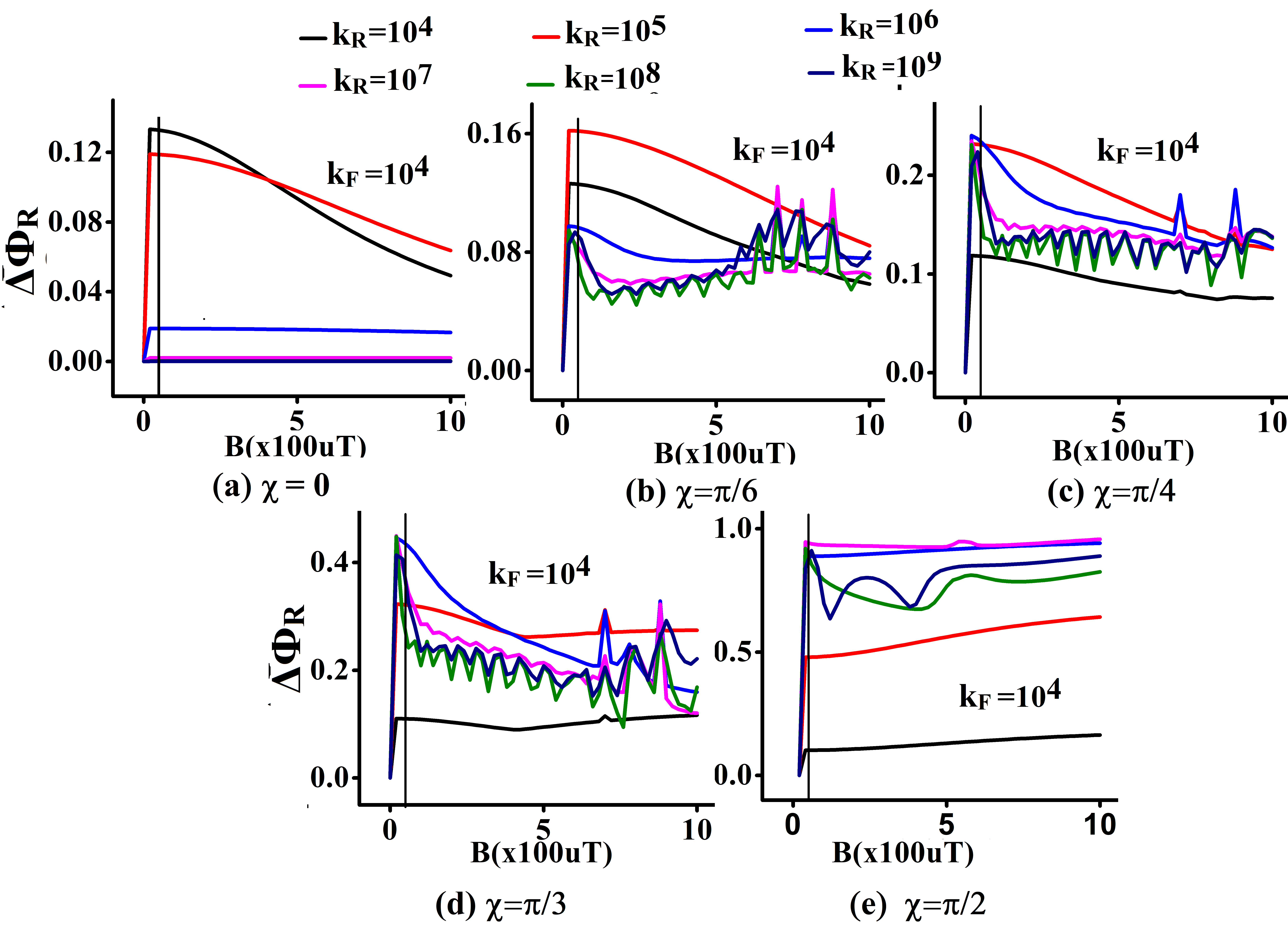}
\caption{Sensitivity ($\triangle\phi_R$) vs $B (\times100\mu T)$ for for five distinct value of $\chi$ showing varying degree of spin selectivity due to CISS ($0^o$, $30^o$, $45^o$, $60^o$, $90^o$). This is shown at  for $k_F=10^4 s^{-1}$ and $k_R\in\lbrace10^4,10^5,10^6,10^7,10^8,10^9\rbrace s^{-1}$. The vertical line marked in each figure correspond to B = 50$\mu T$.}
\label{Functional_Window}
\end{figure}

\begin{figure}[htbp]
\centering
\includegraphics[width=90mm,keepaspectratio]{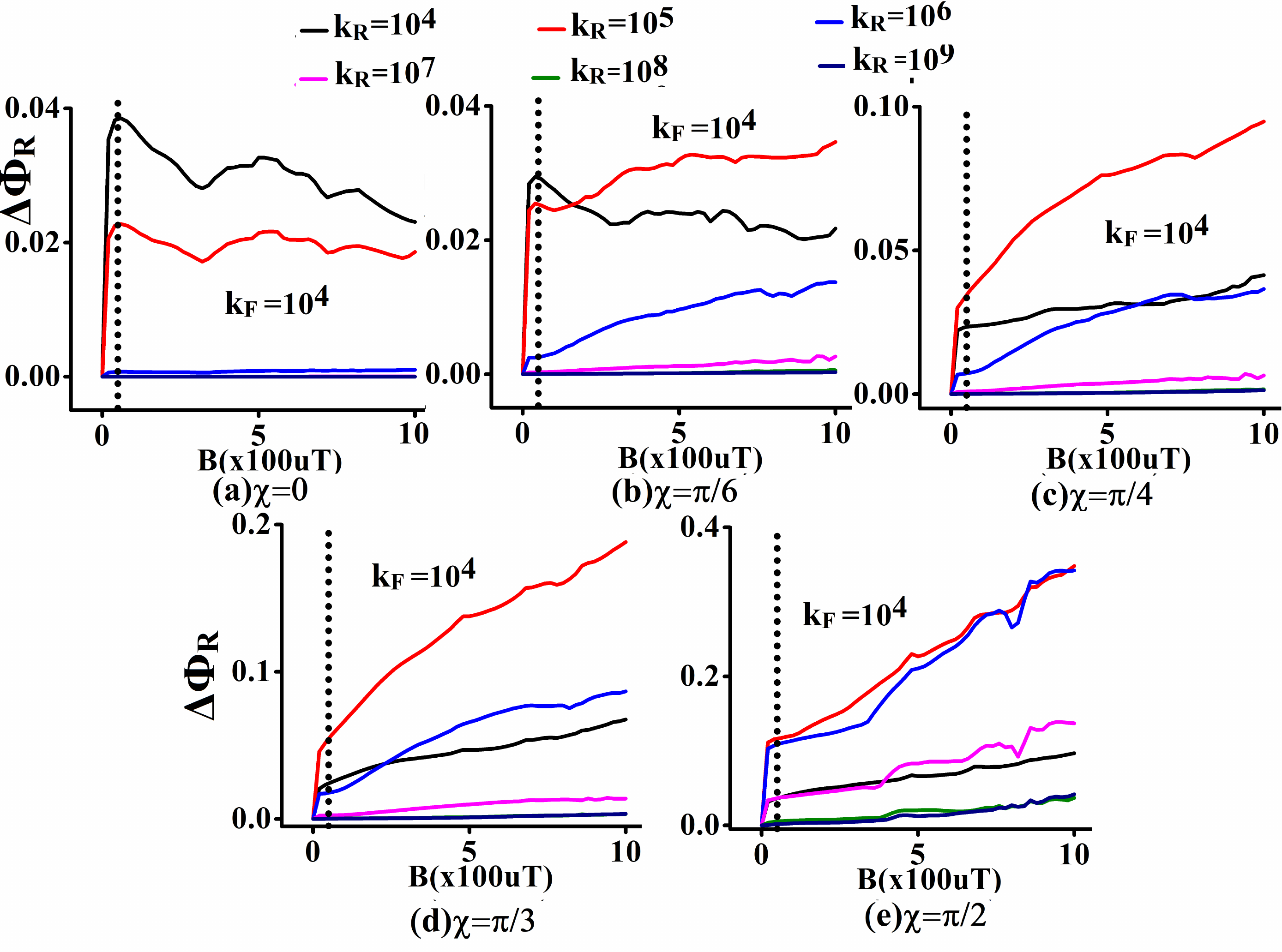}
\caption{Sensitivity ($\triangle\phi_R$) vs $B (\times100\mu T)$ for for five distinct value of $\chi$ showing varying degree of spin selectivity due to CISS ($0^o$, $30^o$, $45^o$, $60^o$, $90^o$). This is shown at  for $k_F=10^4 s^{-1}$ and $k_R\in\lbrace10^4,10^5,10^6,10^7,10^8,10^9\rbrace s^{-1}$ and realistic cryptochrome based RP system. The vertical line marked in each figure correspond to B = 50$\mu T$.}
\label{Functional_WindowPIC}
\end{figure}

We plot functional window in Fig.\ref{Functional_Window} for toy model and in Fig.~\ref{Functional_WindowPIC} for cryptochrome based RP system. We plot  $\triangle\phi_R$ as a function of the external magnetic field. In Fig.\ref{Functional_Window}.a we plot the $\triangle\phi_R$ for $k_F=10^4 s^{-1}$ and $k_R\in\lbrace10^4,10^5,10^6,10^7,10^8,10^9\rbrace s^{-1}$

We observe that sensitivity increases with the channel's spin selectivity $\chi$. For the toy model, the functional window is preserved even with CISS; however, the rate combination at which CISS shows maximum sensitivity and selectivity varies. We observe that the system is showing at CISS $\chi=\frac{\pi}{3}$ for rate combination  $k_F=10^4s^{-1},k_R=10^6s^{-1}$ maximum sensitivity with the functional window feature preserved. At $\chi=\frac{\pi}{2}$ a window is observed to certain degree for combination $k_F=10^4s^{-1},k_R=10^8s^{-1}$ and $k_F=10^4s^{-1},k_R=10^9s^{-1}$ but these values of $k_R$ seems unrealistic based on the timescales of radical pair model as we know.     
A comprehensive figure covering all rate combination for five values of $\chi$ have been given in the Supplementary reading. Here we have picked only $k_F=10^4s^{-1}$. 

For the realistic case (Fig.~\ref{Functional_WindowPIC}), in the functional window, we observe that for $\chi=0$ case, compass selectivity or window is maintained. The introduction of CISS in the avian compass model loses its functional window feature and shows an increase in sensitivity at higher values of the magnetic field. This is not in agreement with behavioral experiments performed on avian compass \cite{wiltschko1972magnetic,winklhofer2013avian,walcott1978animal,ritz2000model}. Thus, the `functional window' could not be modeled even in presence of chiral induced spin selectivity despite an increase in compass sensitivity. This is in general agreement with previous results where functional window is not visible at high reccombination rates~\cite{poonia2017functional}. We are further exploring the CISS based model to see if it can exhibit functional window for some set of compass parameters.

\subsection{\label{ELECTRONELECTRONINTERACTION}Effect of electron-electron Interactions}
So far we have not included the effect of electron-electron interactions.
In this section, we examine the effect of dipolar and exchange interactions along with CISS on compass sensitivity. Both of these effects arise due to the spin property of the electrons. 

Diploar interaction ($D$) is proportional to $r^{-3}$ as given in Eq.~\ref{Dipolar_Interaction} whereas exchange interaction ($J$) is proportional to $e^{-r}$ as given in Eq.~\ref{Exchange_Interaction}~\cite{efimova2008role} where $r$ is the distance between the two electron.
\begin{equation} 
\label{Dipolar_Interaction}
\begin{split}
D(r)=-\frac{3}{2}\frac{\mu_o}{4\pi}\frac{\gamma^2_e\hbar^2}{r^3} \Rightarrow
D(r)/\mu T=-\frac{2.78 \times 10^3}{(r/nm)^3}
\end{split}
\end{equation}
\begin{equation} 
\label{Exchange_Interaction}
\begin{split}
J(r)=J_0 e^{-\beta r}
\end{split}
\end{equation}

When the distance between the two spin r $>$ 1.9nm, we assume in the avian compass that the dipolar Hamiltonian is dominant and the exchange interaction can be neglected \cite{efimova2008role}. In  Eq.\ref{Exchange_Interaction}, $J_0$ and $\beta$ are empirical parameters\cite{jeschke2002determination,o2005influence}. 

\begin{figure}[htbp]
\centering
\includegraphics[width=90mm,keepaspectratio]{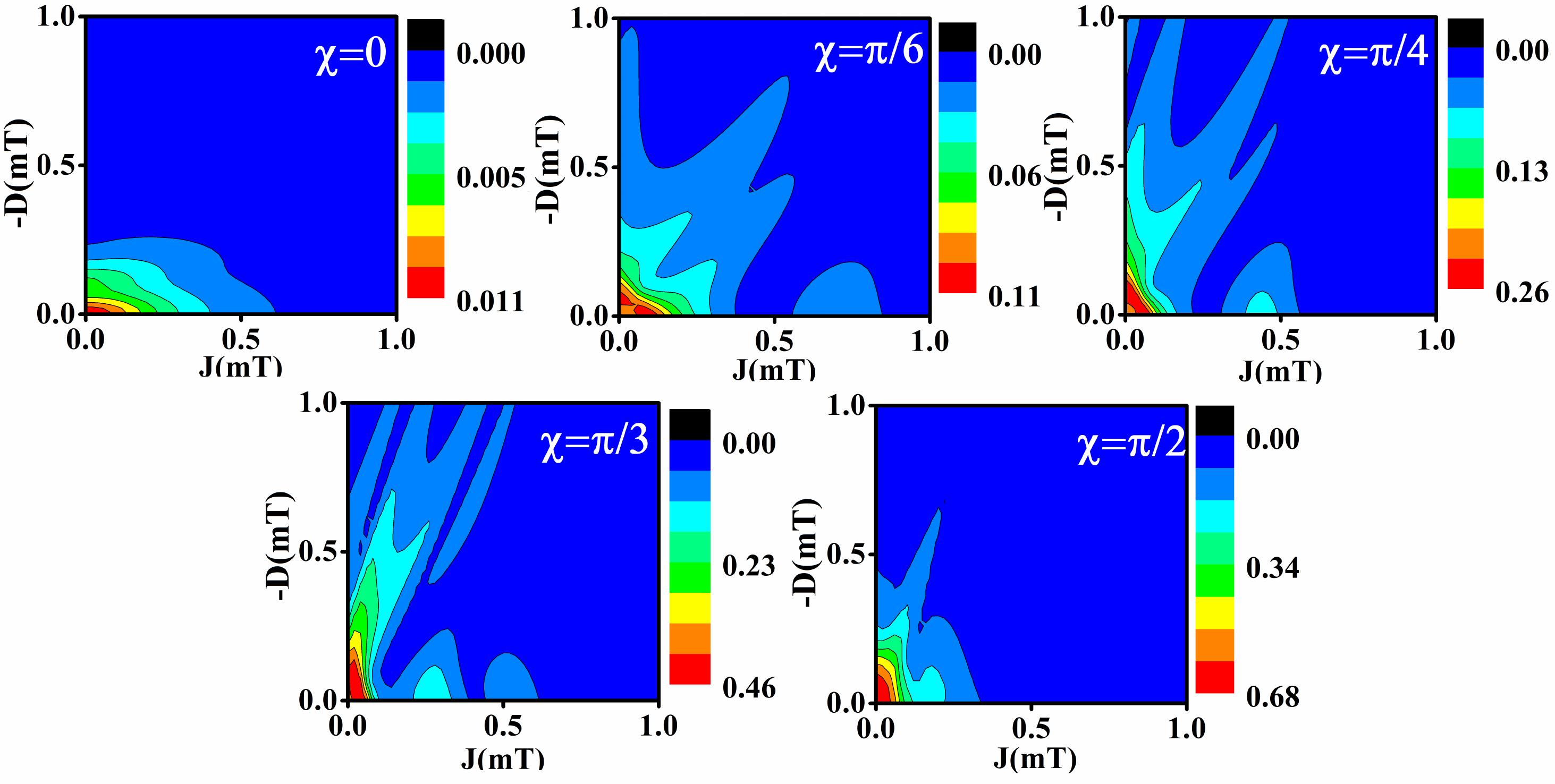}
\caption{Contour plots for sensitivity $\triangle\phi_R$ as a function of $D(mT)$ and $J(mT)$ for five distinct value of $\chi$ showing varying degree of spin selectivity due to CISS ($0^o$, $30^o$, $45^o$, $60^o$, $90^o$) at $k_F=10^6 s^{-1}, k_R=10^8 s^{-1}$.}

\label{High_rateDJ}
\end{figure}

Henceforth, we take the realistic set of rates ($k_F=10^6s^{-1},k_R=10^8s^{-1}$) in our analysis. 
We analyze the contour plots of sensitivity as a function of $D$ and $J$ for five distinct values of $\chi$ (CISS) ($0^o$, $30^o$, $45^o$, $60^o$, $90^o$). Fig.~\ref{High_rateDJ} presents the sensitivity contours for the toy model. In these plots, the dipolar interaction value is taken from -1 mT to 0 and similarly exchange interaction value is considered from 0 to 1mT. For the contour plots, $\triangle\phi_R$ is calculated for $k_F = 10^6 s^{-1}, k_R = 10^8 s^{-1}$.

%

\begin{figure}[htbp]
\centering
\includegraphics[width=90mm,keepaspectratio]{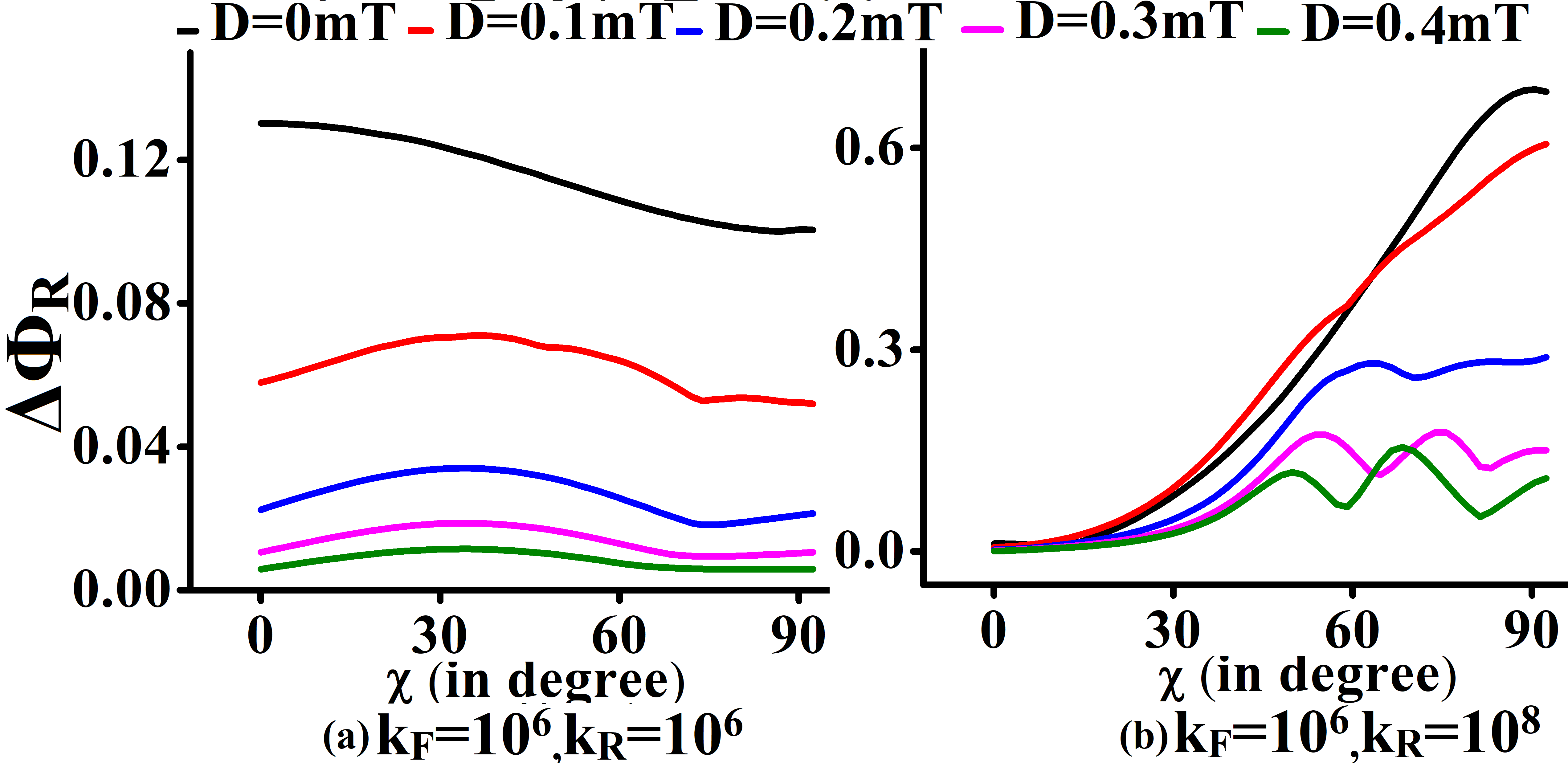}
\caption{Sensitivity ($\triangle\phi_R$) vs $\chi$ for two  combination of signalling state rate $(k_R)$ and ground state rate $(k_F)$  i.e. (a) $k_F = 10^6 s^{-1}, k_R = 10^6 s^{-1}$ (b)$k_F = 10^6 s^{-1}, k_R = 10^8 s^{-1}$ for five values of dipolar interaction when J=0.}

\label{Senstivity_Dipolar}
\end{figure}

In Fig.~\ref{High_rateDJ}, for $\chi=0$ (no CISS), we observe that compass sensitivity is notable for exchange interaction ($J$) from 0 to $\sim $ 0.5mT and dipolar interaction ($D$) from 0 to about $\sim $ -0.2mT. 
As observed in previous results as well, the compass sensitivity increases significantly when CISS is considered (non-zero $\chi$). Interestingly, we observe that for $\chi=\frac{\pi}{6},\frac{\pi}{4}$ and $\frac{\pi}{3}$ the compass sensitivity $\triangle\phi_R$ at D=0.1mT is greater than that at D=0 mT for J=0 mT. This is further confirmed in Fig.~\ref{Senstivity_Dipolar} where the compass sensitivity ($\triangle\phi_R$) is plotted with respect to $\chi$ for five values of dipolar interaction at J=0; This is done for two rate at $k_F = 10^6s^{-1}, k_R = 10^6s^{-1}$ (Fig.~\ref{Senstivity_Dipolar}.(a)) and $k_F = 10^6s^{-1}, k_R = 10^8s^{-1}$ (Fig.~\ref{Senstivity_Dipolar}.(b)). We observe in Fig.~\ref{Senstivity_Dipolar}.(b) that for a certain value of $\chi$, it is not necessary that a higher value of dipolar interaction results in a low value of sensitivity $\triangle\phi_R$. In other words, CISS is canceling the effect of dipolar interaction for some degrees of spin selectivity.

\begin{figure}[htbp]
\centering
\includegraphics[width=90mm,keepaspectratio]{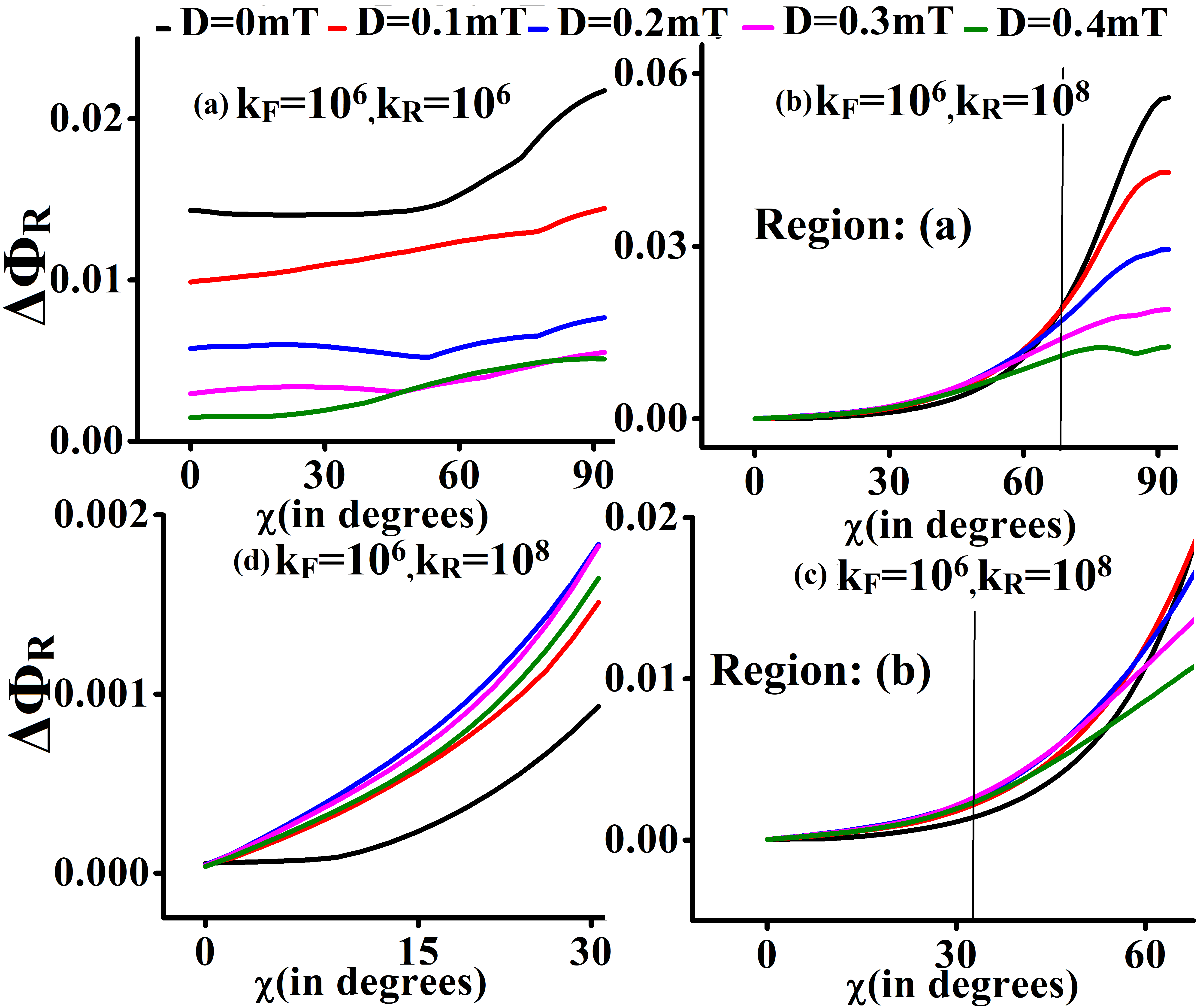}
\caption{[TOP] Sensitivity ($\triangle\phi_R$) vs $\chi$ for two  combination of signalling state rate $(k_R)$ and ground state rate $(k_F)$  i.e. (a)$k_F = 10^6 s^{-1}, k_R = 10^6 s^{-1}$, (b)$k_F = 10^6 s^{-1}, k_R = 10^8 s^{-1}$ for four nuclei cryptochrome system. In (b), region(a) has been enlarged in bottom pane. [BOTTOM] (c) Region (a) is enlargerd and in (d) Region (b) of Fig.(b) has been enlarged showing values of $\chi$ for which $\triangle\phi_R$ at zero dipolar interaction is smaller than the case when dipolar interaction is non-zero.}

\label{Senstivity_Dipolar1}
\end{figure}

To get a more realistic picture, we similarly analyze the cryptochrome based 4-nuclei RP system. In Fig.~\ref{Senstivity_Dipolar1}, we plot the compass sensitivity as a function of $\chi$ for five distinct values of dipolar interaction ($D$).
Fig.~\ref{Senstivity_Dipolar1}(a) and (b) correspond to the plots for $k_F = 10^6 s^{-1}, k_R = 10^6 s^{-1}$ and $k_F = 10^6 s^{-1}, k_R = 10^8 s^{-1}$ respectively. In bottom row we have enlarged region (a) of Fig.~\ref{Senstivity_Dipolar1} (b) for $k_F = 10^6 s^{-1}, k_R = 10^8 s^{-1}$ in Fig.~\ref{Senstivity_Dipolar1} (d), region (b) of Fig.~\ref{Senstivity_Dipolar1} (c) is further enlarged. We observe that a higher value of D does not necessarily mean a lower sensitivity value in a chiral medium. We observe from Fig.~\ref{Senstivity_Dipolar1} (c) that for $\chi \leq 50^o$, we observe that the sensitivity is higher in presence of dipolar interaction due to CISS.

\begin{figure}[t]
\centering
\includegraphics[width=90mm,keepaspectratio]{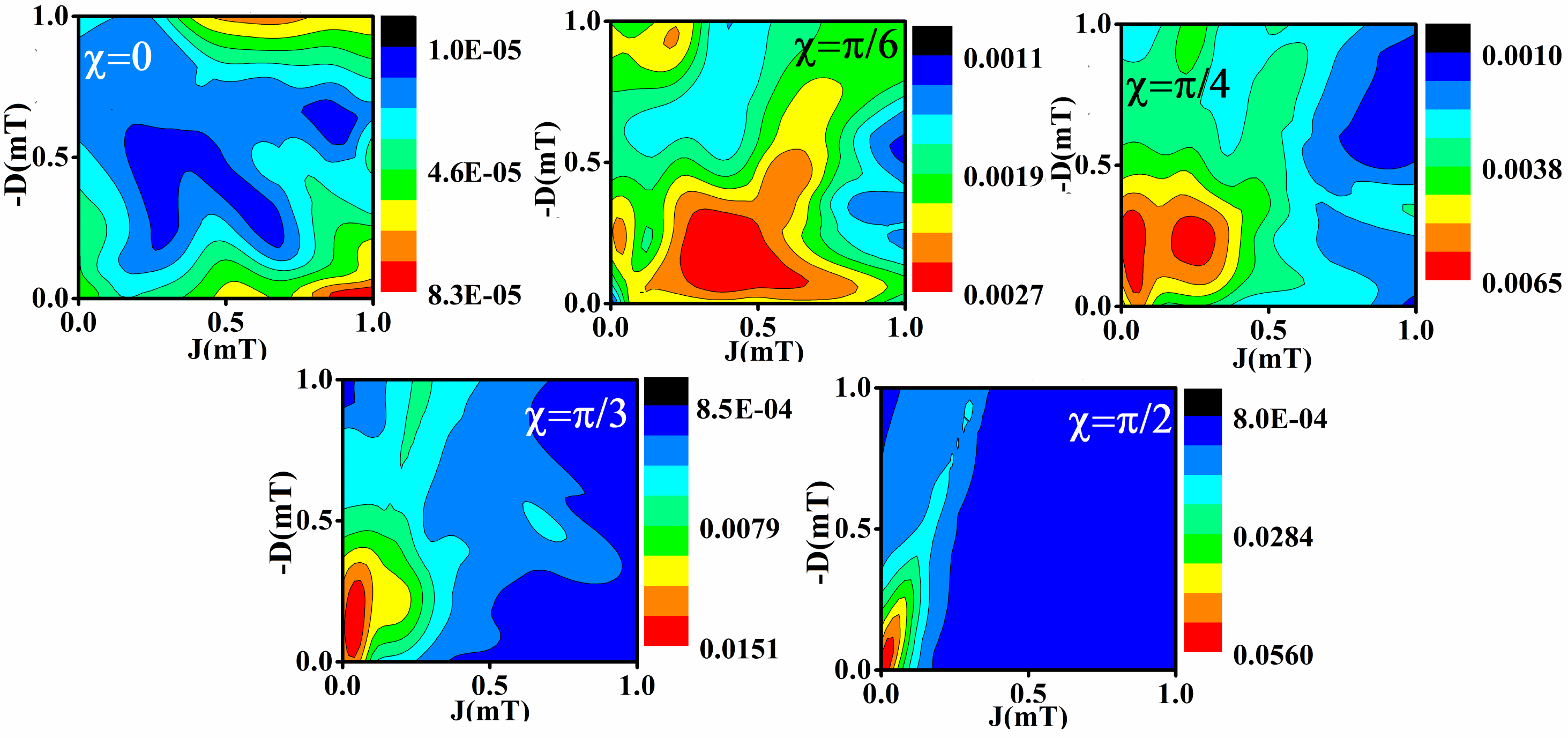}
\caption{Contour plots for sensitivity ($\triangle\phi_R$) as a function of $D(mT)$ and $J(mT)$ for five distinct value of $\chi$ showing varying degree of spin selectivity due to CISS ($0^o$, $30^o$, $45^o$, $60^o$, $90^o$) at $k_F=10^6s^{-1}, k_R=10^8s^{-1}$ for 4-nuclei $FAD_{2N}-TrpH_{2N}$ based RP system.}
\label{Senstivity_DipolarJ1}
\end{figure}

Further, we analyze the effect of exchange interaction in conjunction with dipolar interaction along with CISS for 4-nuclei RP system, as done earlier for the toy model in Fig.~\ref{High_rateDJ}. In Fig.~\ref{Senstivity_DipolarJ1}, we plot the compass sensitivity ($\triangle\phi_R$) as function of $D$ and $J$ at rate $k_F = 10^6 s^{-1}, k_R = 10^8 s^{-1}$ for five distinct values of $\chi$ ($0^o$, $30^o$, $45^o$, $60^o$, $90^o$).

We found that chirality affects the dipolar and exchange value for which we attain the maxima of compass sensitivity. When $\chi$=0, we observe that maximum sensitivity is no longer at $J=0$ and $D=0$.  We observe the compass sensitivity maxima in region where $J\neq0,D\neq0$ for $\chi=\frac{\pi}{6}$ and $\chi=\frac{\pi}{4}$. This region shrinks and again converges to origin ($J=0$ and $D=0$) as $\chi$ is further increased. Full CISS results in a small range of dipolar and exchange values with compass sensitivity showing maxima near origin. This confirms our result of Fig.\ref{Senstivity_Dipolar1} where we showed for $\chi\leq50^0$ dipolar interaction is required for increased sensitivity. In addition to this for $\chi=\frac{\pi}{6}$ and $\chi=\frac{\pi}{4}$, we find finite J also result in increase sensitivity. Thus CISS cancels the deleterious effect of dipolar and exchange interactions in these regimes. For cases $\chi=\frac{\pi}{3}$ and $\chi=\frac{\pi}{2}$ , however, we observe that a lower value of D and J corresponds to the increased sensitivity. 

Finally, we have provided similar contour plots for toy model and 4-nuclei based RP system at $k_F = 10^6 s^{-1}, k_R = 10^6 s^{-1}$ in the Appendix. There, we observe that we get a lesser range of $D$ and $J$ values for which we get significant compass sensitivity in the toy model. For 4-nuclei RP system, we do observe sensitivity rise around J$\sim$ 0.4mT, D$\sim$1mT when $\chi=0$. An increase in sensitivity was observed with increase in CISS. A larger range of J$\in\lbrace0,0.5mT\rbrace$ and D$\in\lbrace0,1mT\rbrace$ shows us an increase in sensitivity as $\chi$ increases from 0 to $\frac{\pi}{2}$.
Hence, in a nutshell, CISS allows us to have sensitivity at higher recombination rates, and at these higher rates we observe mitigation of detrimental effects of dipolar and exchange interactions to a certain extent.

\section{Conclusion}
In conclusion, we observe that the compass sensitivity is enhanced by chiral induced spin selectivity (CISS) effect for realistic recombination rate values. The functional window characteristics of the compass is, however, lost. We also observe that dipolar and exchange interaction are generally detrimental in the functioning of the avian compass, but it can be mitigated to a certain extent by CISS. In general, CISS seems to favor the RP model by enhancing its parameter regime to more realistic values. In future work, we plan to delve more deeply on the parameter regions where CISS is canceling the deleterious effects of dipolar and exchange interaction. We also plan to examine RF disruption property of avian compass in presence of CISS. Finally, we would like to do this study on the full cryptochrome based RP system that would require larger computational resources.
\begin{acknowledgments}
The author would like to thank Jiate Luo, Department of Chemistry, University of Oxford, UK and Peter J. Hore, Department of Chemistry, University of Oxford, UK for their invaluable and insightful communication. This work is supported by the Science and Engineering Research Board, Department of Science and Technology (DST), India with Grant No. CRG/2021/007060. The authors would also like to thank Department of Electronics and Communication, IIT Roorkee and Ministry of Education, Government of India for supporting Y.T.'s graduate research.
\end{acknowledgments}

\appendix
\section{Appendices}
\begin{figure}[htbp]
\centering
\includegraphics[width=90mm,keepaspectratio]{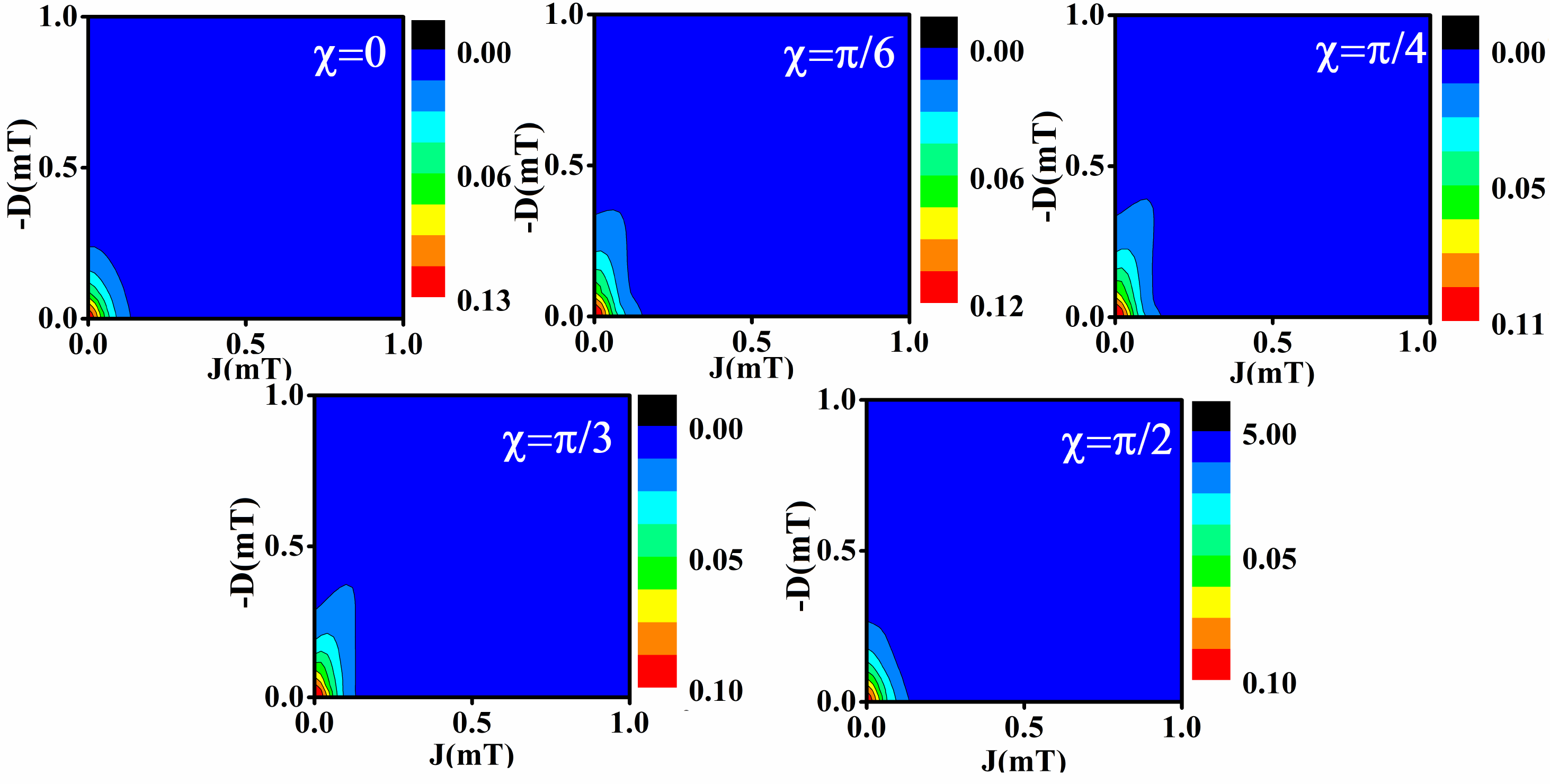}
\caption{Contour plots for sensitivity ($\triangle\phi_R$) as a function of $D(mT)$ and $J(mT)$ for five distinct value of $\chi$ showing varying degree of spin selectivity due to CISS ($0^o$, $30^o$, $45^o$, $60^o$, $90^o$) at $k_F=10^6 s^{-1}, k_R=10^6 s^{-1}$ for the toy model.}
\label{Low_rateDJ}
\end{figure}

Fig.\ref{Low_rateDJ}, is the contour plot of senstivity with respect to D and J for toy model at $k_F=10^6s^{-1}, k_R=10^6s^{-1}$. At D=0 and J=0, we observe the maximum value of sensitivity, and it reduces if we move in any direction from that point. Here we observe that exchange and dipolar interaction have a maximum limit beyond which the sensitivity change is negligible. The dipolar interaction is limited at $\sim \pm$ 0.4mT (for $\chi=\frac{\pi}{4}$) and exchange interaction at $\sim $ 0.1mT. We also observe that CISS affects this limit, but the change in limit is not that much.

\begin{figure}[htbp]
\centering
\includegraphics[width=90mm,keepaspectratio]{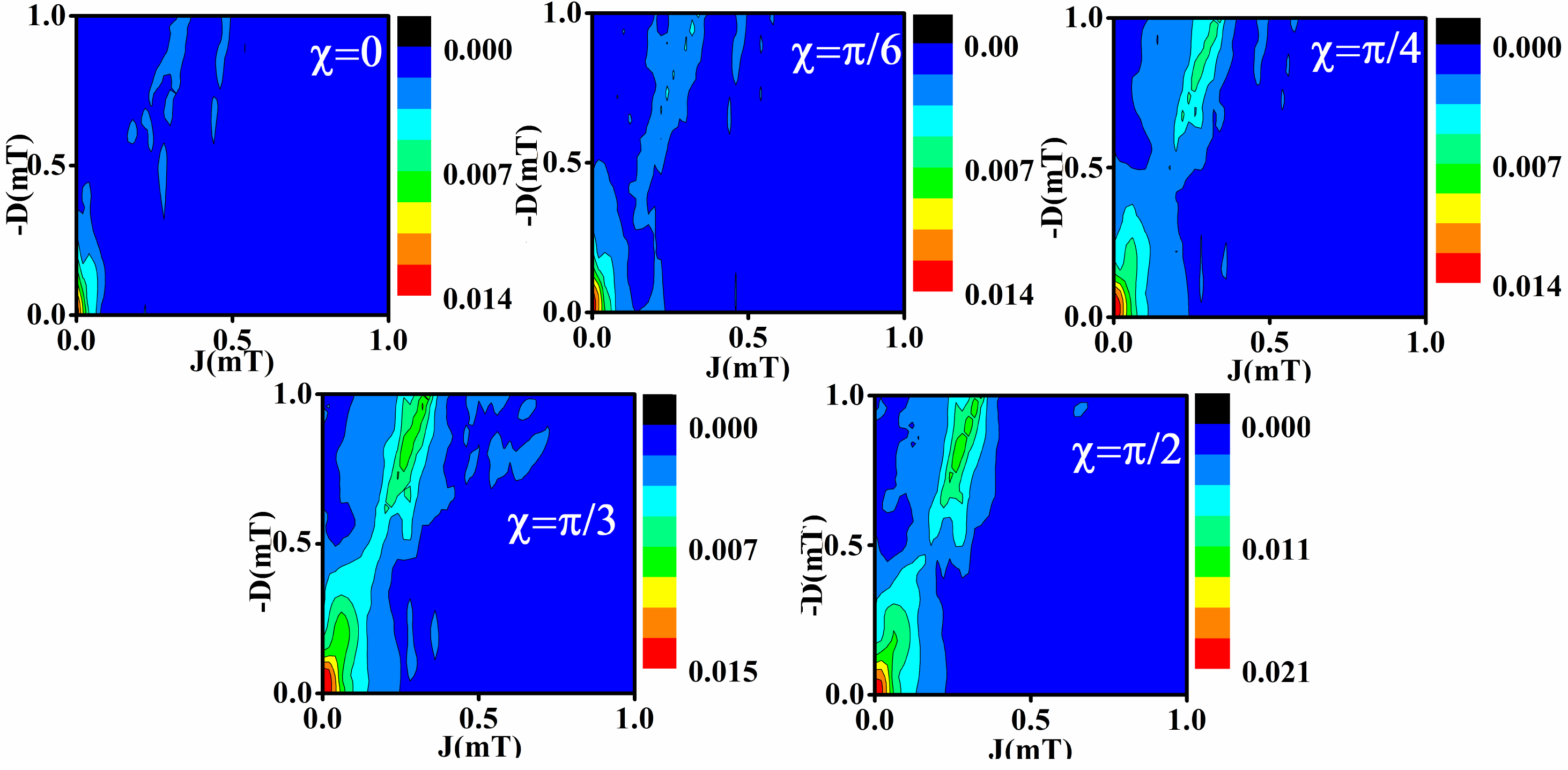}
\caption{Contour plots for sensitivity $\triangle\phi_R$ as a function of $D(mT)$ and $J(mT)$ for five distinct value of $\chi$ showing varying degree of spin selectivity due to CISS ($0^o$, $30^o$, $45^o$, $60^o$, $90^o$) at $k_F=10^6 s^{-1}, k_R=10^6 s^{-1}$ for 4-nuclei $FAD_{2N}-TrpH_{2N}$ based RP system.}
\label{Senstivity_DipolarJ2}
\end{figure}
Fig.\ref{Senstivity_DipolarJ2}, is the contour plot of sensitivity with respect to D and J for four nuclei taken from cryptochrome radical pair model at $k_F=10^6s^{-1}, k_R=10^6s^{-1}$. For $\chi=0$, we observe that a higher value of dipolar interaction gives us some sensitivity, although the value is small for a finite value of J. This is in agreement with the result reported in \cite{efimova2008role}, where partial cancellation of dipolar interaction occurs due to finite value of exchange interaction. However, it is interesting to note that for $\chi=\frac{\pi}{6},\frac{\pi}{4},\frac{\pi}{3}$ and $\frac{\pi}{2}$, we observe that sensitivity increases manifold as well as the larger combination of J and D gives us sensitivity. Hence this J/D cancellation increases as the system selectivity increase due to chirality.

\bibliography{apssamp}

\end{document}